\documentclass[preprint,aps,eqsecnum,graphicx]{revtex}

\def\beq#1{\begin{equation} \label{#1}}
\def\eeq{\end{equation}}
\def\bra#1{\left\langle #1\right\vert}
\def\ket#1{\left\vert #1\right\rangle}

\begin{document}
{
\tighten
\begin{center}
{\Large\bf New systematics in charmless strange $B^+ \rightarrow VP$ decays}  \\
\vrule height 2.5ex depth 0ex width 0pt
\vskip0.8cm
Harry J. Lipkin\,$^{b,c}$\footnote{e-mail: \tt ftlipkin@weizmann.ac.il} \\
\vskip0.8cm
{\it
$^b\;$School of Physics and Astronomy \\
Raymond and Beverly Sackler Faculty of Exact Sciences \\
Tel Aviv University, Tel Aviv, Israel\\
\vbox{\vskip 0.0truecm}
$^c\;$Department of Particle Physics \\
Weizmann Institute of Science, Rehovot 76100, Israel \\
and\\
High Energy Physics Division, Argonne National Laboratory \\
Argonne, IL 60439-4815, USA
} 
\end{center}

\vspace*{0.8cm}
\centerline{\bf Abstract}
\vspace*{4mm}

Latest data on charmless strange vector-pseudoscalar $B^+$ decays now including
$B^+\rightarrow \rho^+ K^o$ confirm a simple penguin model in which the gluon
$G$ in an initial $\bar s u G$ state fragments equally into $u \bar u$, $d \bar
d$ and $s \bar s$ and all form factors are equal. A search for possible 
additional contributions shows only a few signals not obscured by experimental
errors whose implications are discussed. The experimental value of $0.25 \pm
0.11$ for the ratio of the branching ratios  $BR(B^+ \rightarrow  K^{*+}\eta)$
to $BR(B^+\rightarrow K^{*+}\eta')$ confirms  the parity selection rule
prediction $0.32$. Large violations arise in a  new sum  rule for the sum of
these branching ratios, analogous to the similar pseudoscalar sum rule
including $K^+\eta$ and $K^+\eta'$. Indications for either an electroweak
penguin contribution 
or additional admixtures like instrinsic charm in the $\eta-\eta'$
system remain to be clarified.  An alternative
symmetry description with new predictive power clarifies the simple penguin 
approximation and presents new predictions which can be tested experimentally.
The fragmentation of the $\bar s u G$ state into two mesons is
described by a strong interaction S-matrix dominated by nonexotic  hadron
resonances in  multiparticle intermediate states. 

\vfill\eject
\section{Implications of experimental systematics in charmless strange 
$B^+$ vector-pseudoscalar decays}
    
New experimental data \cite{bko} on 
$B^+\rightarrow \rho^+ K^o$ 
satisfy the prediction\cite{bkpfsifin,HJLCharm,PKETA}
\beq{bkrpred}
BR(B^+\rightarrow \rho^+ K^o) = BR(B^+\rightarrow \phi K^+) 
\end{equation}
The paper\cite{bko} quotes the assumption\cite{grorosbvp,chiang,chiangros}   
$p'_V=-p'_P$ where $p'_V$ and $p'_P$
denote the amplitudes for the spectator quark to appear in the vector or
pseudoscalar meson. The relation between the magnitudes 
\beq{formfac}
|p'_V|=|p'_P|
\end{equation}
is actually sufficient to obtain the prediction (\ref{bkrpred}).

This neglect of differences between vector and pseudoscalar form factors at the
weak and spectator vertices appears completely unjustified in the conventional
descriptions. We therefore look for an alternative approach and search for 
some underlying symmetry.   A penguin diagram for $B^+$ decays into charmless
strange vector-pseudoscalar states begins with a weak interaction that produces
an $\bar s u$ gluon state.  The gluon and $\bar s$ are emitted in opposite
directions in the rest system of the spectator quark.   This finishes the weak
part of the process.  The gluon must then interact with both others to produce
the final state. The common assumption that the gluon first produces a $q \bar
q$ pair before interaction with the others can be questioned. We look for other 
descriptions that go beyond this approximation.
    If the strong interaction conserves flavor SU(3) symmetry, the
symmetry alone places serious constraints on the observable branching
ratios. These are completely independent of the detailed dynamics and
approximations like factorization used in conventional approaches. We shall show
that this symmetry approach leads to the relation (\ref{formfac}) without any
discussion of form factors.  

    An $\bar s u$ gluon state is an octet in flavor SU(3) and is a vector in
the $V$-spin subgroup of SU(3) with $V_z = 1$.  Its eigenvalue can be either
even or odd under $G_V$ parity, the analog of $G$ parity with isospin rotated
into $V$ spin. Since $G_V$ is conserved  in QCD interactions the amplitude for
the final two meson state is a linear combination of an even $G_V$ amplitude
and an odd $G_V$ amplitude. If both amplitudes contribute to the final
two-meson state the relative branching ratios will be unpredictable. The
branching ratios depend upon the relative magnitude and phase of the two
amplitudes which are determined by unknown QCD dynamics. If only the odd $G_V$
amplitude contributes, the  relation between (\ref{formfac}) immediately
follows. The odd $G_V$ amplitude has the flavor quantum numbers of the kaon and
leads to  branching ratios determined from $V$ spin like those from decay of a
high mass  kaon.

    The search for a justification for neglecting the other even $G_V$
amplitude is discussed in detail below. It follows automatically if  the QCD
strong interaction S-matrix is dominated by S channel resonances which all have
odd GV parity.

We now examine the systematics of the experimental data and then discuss the
symmetry picture in detail including new predictions for other final states.

\subsection{Experimental tests of the simplest extreme penguin model}

We first examine the simplest gluonic 
penguin model in which the gluon
$G$ in an initial $\bar s u G$ states fragments equally into $u \bar u$, $d
\bar d$ and $s \bar s$, all form factors are equal and the OZI 
rule\cite{ALS,PATOZ,Okubo,Zweig,Iizuka} is
respected; i.e. the $q \bar q$ pair produced by the gluon does not end in the same
final meson. This gives the relations
\beq{nonexot}
\begin{array}{ccccccc}
2BR(B^+\rightarrow K^+ \omega) &=2BR(B^+\rightarrow K^+ \rho^o) &=
BR(B^+\rightarrow K^o \rho^+) =
\\
=2BR(B^+\rightarrow K^{*+} \pi^o) &=BR(B^+\rightarrow K^{*o} \pi^+) &=
BR(B^+\rightarrow \phi K^+) 
\\ \rm{Data}&\rm{From}&\rm{BaBar}
\\
12.2 \pm 1.2 \pm 0.8 &= 10.2\pm 1.6 \pm 1.6 &=  8.0\pm 1.4 \pm 0.5 =
\\
 = 13.8 \pm 4.0 \pm 2.6 &=  13.5 \pm 1.2 \pm 0.9 &=  8.4 \pm 0.7 \pm 0.7
\\ \rm{Data}&\rm{From}&\rm{HFAG-Avg}
\\
13.6 \pm 1.0  &= 8.5\pm 1.1 &= 8.0\pm 1.5  =
\\
 = 13.8 \pm 4.6 &=  10.7 \pm 0.8 &= 8.3 \pm 0.65
\end{array}
\end {equation}
where we have used the data from BaBar quoted by HFAG\cite{HFAG}  and also the 
HFAG Average listed below in units of $10^{-6}$ and do not include the final
states including the $\eta$ and $\eta'$. This is still reasonably good and not
sufficiently precise to pinpoint other contributions omitted in this simple
picture. The one exception is the large $BR(B^+\rightarrow K^+ \omega)$ seen
in the HFAG average although not in the BaBar data. 

Each of the two individual lines of equalities in eq.(\ref{nonexot}) is
independent of the form factor assumption, (\ref{formfac}). The first line
gives equalities between transitions to final states where the vector meson
contains the spectator quark; the second gives equalities between transitions 
to final states where the vector meson contains the $\bar s$ antiquark from the
weak vertex. The assumption of equal form factors (\ref{formfac}) is needed only
to relate the two lines. The data and in particular the relation
(\ref{bkrpred}) confirm the equality between the two lines and therefore the 
assumption (\ref{formfac}).

The relations for the neutral decays corresponding to (\ref{nonexot})
for the charged decays are:

\beq{nonexoto}
\begin{array}{ccccccc}
2BR(B^o\rightarrow K^o \omega) &=2BR(B^o\rightarrow K^o \rho^o) &=
BR(B^o\rightarrow K^+ \rho^-) =
\\
=2BR(B^o\rightarrow K^{*o} \pi^o) &=BR(B^o\rightarrow K^{*+} \pi^-) &=
BR(B^o\rightarrow \phi K^o) 
\\ \rm{Data}&\rm{From}&\rm{HFAG-Avg}
\\
10.4 \pm 1.4  &= 10.8\pm 2.0 & = 15.3\pm 3.7 
\\
 = 0.0 \pm 2.6 &= 9.8 \pm 1.1  &= 8.3 \pm 1.2
\end{array}
\end {equation}

These are also in reasonable agreement except for the small 
$B^o\rightarrow K^{*o} \pi^o$ decay.

The relations (\ref{nonexot}) are also obtainable by noting that initial state 
has the flavor and parity quantum numbers of a kaon and using SU(3) flavor
symmetry to relate the decays of a high mass kaon. 

\subsection{Experimental data used in our analyses}

We have used the BaBar experimental data\cite{HFAG} and also the 
HFAG Average listed below in
units of $10^{-6}$ and leave the combination of statistical and systematic
errors for the reader.

\beq{pengoBRexp2}
\begin{array}{ccccccc}
\rm{Transition} & &\rm{Babar ~ Data } & &
{\rm  HFAG-Avg }& &{\rm Momentum} \\
BR(B^+\rightarrow K^+ \omega) &=& 6.1 \pm 0.6 \pm 0.4
 &=& 6.8 \pm 0.5 ; &  &{\rm p} = 2557\\
BR(B^+\rightarrow K^+ \rho^o) &=& 5.1\pm 0.8 \pm 0.8
 &=& 4.25\pm 0.56 ; & &{\rm p} =  2558\\
BR(B^+\rightarrow K^o \rho^+) &=& 8.0\pm 1.4 \pm 0.5
 &=& 8.0\pm 1.5 ; & & {\rm p} = 2558\\
BR(B^+\rightarrow K^{*+} \eta) &=& 18.9 \pm 1.8 \pm 1.3
 &=& 19.3 \pm 1.6; & &{\rm p} =  2534 \\
BR(B^+\rightarrow K^{*+} \eta')&=& 4.9 \pm 1.9 \pm 0.8
 &=& 4.9 \pm 2.1 ; & &{\rm p} =  2472\\
BR(B^+\rightarrow K^{*o} \pi^+) &=& 13.5 \pm 1.2 \pm 0.9
 &=& 10.7 \pm 0.8; &  &{\rm p} = 2562\\
BR(B^+\rightarrow K^{*+} \pi^o) &=& 6.9 \pm 2.0 \pm 1.3
 &=& 6.9 \pm 2.3 ; & &{\rm p} =  2562\\
BR(B^+\rightarrow \phi K^+) &=& 8.4 \pm 0.7 \pm 0.7
 &=& 8.3 \pm 0.65 ; & &{\rm p} =  2516
\end{array}
\end {equation}
 
\beq{pengoBRexp0}
\begin{array}{ccccccc}
\rm{Transition} & &\rm{Babar ~ Data } & &
{\rm  HFAG-Avg }& &{\rm Momentum} \\
BR(B^o\rightarrow K^o \omega) &=& 6.2 \pm 1.0 \pm 0.4
 &=& 5.2 \pm 0.7 ; &  &{\rm p} = 2557\\
BR(B^o\rightarrow K^o \rho^o) &=& 4.9\pm 0.8 \pm 0.9
 &=& 5.4\pm 1.0 ; & &{\rm p} =  2558\\
BR(B^o\rightarrow K^+ \rho^-) &=& 
 &=& 15.3\pm 3.7 ; & & {\rm p} = 2558\\
BR(B^o\rightarrow K^{*o} \eta) &=& 16.5 \pm 1.1 \pm 0.8
 &=& 15.9 \pm 1; & &{\rm p} =  2534 \\
BR(B^o\rightarrow K^{*o} \eta')&=& 3.8 \pm 1.1 \pm 0.5
 &=& 3.8 \pm 1.2 ; & &{\rm p} =  2472\\
BR(B^o\rightarrow K^{*o} \pi^o) &=& 
 &=& 0.0 \pm 1.3; &  &{\rm p} = 2562\\
BR(B^o\rightarrow K^{*+} \pi^-) &=& 11.0 \pm 0.4 \pm 0.7
 &=& 9.8 \pm 1.1 ; & &{\rm p} =  2562\\
BR(B^o\rightarrow \phi K^o) &=& 8.4 \pm 1.5 \pm 0.5
 &=& 8.3 \pm 1.2 ; & &{\rm p} =  2516
\end{array}
\end {equation}

\subsection{Possible signals from violations of the simplest gluonic penguin
model}

Our approach here is complementary to the extensive analysis \cite{chiang}
which uses a more detailed model with more parameters to fit much more data. We
concentrate here on the simple penguin model which seems to do too well and
look for the inevitable signals for its breakdown which are mainly still
obscured by experimental errors. We include updated data for  $B^o\rightarrow
K^{*o} \pi^o$, $B^o\rightarrow K^o \rho^o$ and $B^+\rightarrow K^o \rho^+$ not
yet available to  ref.\cite{chiang} and pointed out there as ``soon to be
seen".

We first note that the penguin diagram produces an isospin eigenstate with
$I=1/2$. Thus the transtions to the two $K^*\pi$ and the two $K\rho$ final
states are related by isospin and completely independent of all form factors.
The factors of 2 appearing  in  eqs. (\ref{nonexot}) and (\ref{nonexoto}) are
isospin Clebsch-Gordan coefficients. Any violation of these isospin equalities
indicates either an isospin violation, as in an electroweak penguin
contribution, or an isospin $I=3/2$ contribution, as in a tree diagram
contribution. The two violations of the simple gluonic penguin that we have
noted above, the large  $BR(B^+\rightarrow K^+ \omega)$ and the small 
$BR(B^o\rightarrow K^{*o}\pi^o)$ have been disucssed in ref.\cite{chiang} and 
can be due to electroweak penguin contributions. 

The relation \beq{krhomeg} \frac{BR(B^+\rightarrow K^+
\omega)}{BR(B^+\rightarrow K^+ \rho^o)} = 1 \end{equation} follows from any
combination of penguin and tree amplitudes\cite{PKETA} but can be broken by an
electroweak penguin\cite{chiang}. The experimental data in ref.\cite{chiang}
gave $1.3 \pm 0.3$ consistent with 1. The new data violate the relation
(\ref{krhomeg}) and may indicate an EWP contribution. The small
$BR(B^o\rightarrow K^{*o} \pi^o)$ is predicted in ref.\cite{chiang} as due to
the EWP contribution. Their prediction of $1 \times 10^{-6}$ is still 
consistent with the smaller value in the new data. 

\subsection{A good relation between the ratio of 
$BR(B^+\rightarrow K^{*+} \eta')$ to $BR(B^+\rightarrow K^{*+} \eta)$}

The high mass kaon model also predicts the observed suppression of  
$BR(B^+\rightarrow K^{*+} \eta')$ relative to $BR(B^+\rightarrow K^{*+} \eta)$.
A high mass $K^+$ goes into a
high mass $\pi^+$ under the $U$-spin reflection $s\leftrightarrow d$
 The decay $\pi^+ \rightarrow \rho^+ \pi^o$ is allowed by  $G$
parity; the decay $\pi^+ \rightarrow \rho^+ \eta$ is forbidden. $U$-spin
reflections of these decays change:
\beq{ureflec}
\pi^+\leftrightarrow K^+; ~ ~ ~
\rho^+\leftrightarrow K^{*+}; ~ ~ ~ G\leftrightarrow G_V; ~ ~ ~
\eta_u \pm \eta_d \leftrightarrow \eta_u \pm \eta_s
\end{equation}
where 
$G_V$ parity is defined with the $V$-spin (us)
subgroup of flavor SU(3) like the
ordinary G parity is defined with isospin and 
the pseudoscalar flavor states $\ket{\eta_i}$ are defined as pseudoscalars
created from  a $q\bar q$ pair with flavors $i$ that can be $u$, $d$ or
$s$ and the nonstrange pseudoscalars are 
\beq{etanon}
\ket{\pi^o} \equiv \frac{\ket {\eta_u} - \ket {\eta_d}}{\sqrt {2}}; ~ ~ ~
\ket{\eta_n} \equiv \frac{\ket {\eta_u} + \ket {\eta_d}}{\sqrt {2}}
\end {equation}

These show that the decay 
$K^+ \rightarrow  K^{*+} (\eta_u - \eta_s) $ is allowed by  $G_V$ parity;
while the decay $K^+ \rightarrow K^{*+}  (\eta_u + \eta_s)$ is forbidden.
Since the  $\eta$ and $\eta'$ wave functions in the standard mixing model
combines the $\eta_u$ and $\eta_s$ components with a positive phase in the
$\eta'$ and a negative phase in the $\eta$, the transition matrices in the 
$\eta-\eta'$ basis are related as
\beq{parselvp}
\frac{\bra{K^{*+}  \eta'}T\ket{B^+}}{\bra{K^{*+}  \eta}T\ket{B^+}}=
\frac{\bra{\eta'} \eta_u - \eta_s\rangle}{\bra{\eta} \eta_u - \eta_s\rangle}
= -\frac{1}{\sqrt 8} =
\frac{\bra{\eta'} \eta_d - \eta_s\rangle}{\bra{\eta} \eta_d - \eta_s\rangle}=
\frac{\bra{K^{*o}  \eta'}T\ket{B^+}}{\bra{K^{*o}  \eta}T\ket{B^+}}
\end{equation}
where we have used isospin and the fact the the penguin final state is
isoscalar to include the neutral decays and substituted the mixing angle
commonly used\cite{Bramon}

\beq{etamixram}
\ket{\eta} = \frac{\ket{\eta_u}+\ket{\eta_d}-\ket{\eta_s}}{\sqrt 3}; ~ ~ ~
\ket{\eta'} = \frac{\ket{\eta_u}+\ket{\eta_d}+2\ket{\eta_s}}{\sqrt 6}
\end{equation} 

This qualitatively predicts the observed suppression of $BR(B^+\rightarrow
K^{*+} \eta')$ relative to $BR(B^+\rightarrow K^{*+} \eta)$ and notes correctly
that the suppression factor for the vector-pseudoscalar  case is much less 
than  the infinite suppression for the two pseudoscalar case predicted by this
mixing (\ref{etamix}). 
\beq{parselpp}
\frac{\bra{K^+  \eta}T\ket{B^+}}{\bra{K^+  \eta'}T\ket{B^+}}=
\frac{\bra{\eta} \eta_u + \eta_s\rangle}{\bra{\eta'} \eta_u + \eta_s\rangle}
= 0
\end{equation}
For a better approximation we use a general mixing angle
\beq{etamix}
\ket{\eta'} =\ket{\eta_n} \cos \theta + \ket{\eta_s} \sin \theta; ~ ~ ~
\ket{\eta} = \ket{\eta_n} \sin \theta - \ket{\eta_s} \cos \theta 
\end{equation} 
Then 
\beq{ratpp}
\frac{\bra{K^+  \eta}T\ket{B^+}}{\bra{K^+  \eta'}T\ket{B^+}}=
\frac{\bra{\eta} \eta_u + \eta_s\rangle}{\bra{\eta'} \eta_u + \eta_s\rangle}
=\frac{\sin  \theta -\sqrt{2} \cos\theta}{\cos \theta + \sqrt{2}\sin \theta}
=\frac{\tan \theta - \sqrt{2}}{ 1 +\sqrt{2} \tan \theta}
\end{equation}
Solving for the mixing angle $\theta$ in terms of the experimentally measured
ratio denoted by $X$ gives
\beq{tantheta}
\tan \theta = \frac{\sqrt{2} +X}{1-X \sqrt{2}}; ~ ~ ~ X \equiv 
\frac{\bra{K^+  \eta}T\ket{B^+}}{\bra{K^+  \eta'}T\ket{B^+}}
\approx \sqrt{\left[\frac{2.2 \pm 0.3}{69.7 \pm 3.8}\right]} = .178 
\end{equation}
This then predicts
\beq{ratvp}
\frac{\bra{K^{*+}  \eta'}T\ket{B^+}}{\bra{K^{*+}  \eta}T\ket{B^+}}=
\frac{\bra{\eta'} \eta_u - \eta_s\rangle}{\bra{\eta} \eta_u - \eta_s\rangle}
=\frac{ 1 -\sqrt{2} \tan \theta}{\tan \theta + \sqrt{2}}
=\frac{1+2X\sqrt{2}}{X-2\sqrt{2}} = -.567
\end{equation}

or
\beq{bran}
\frac {BR(B^+\rightarrow K^{*+} \eta')}{BR(B^+\rightarrow K^{*+} \eta)}
=\frac {BR(B^o\rightarrow K^{*o} \eta')}{BR(B^o\rightarrow K^{*o} \eta)}
\approx \left| \frac{\bra{K^{*+}\eta'}T\ket{B^+}}
{\bra{K^{*+}  \eta}T\ket{B^+}}\right|^2 = 0.32
\end{equation}
The experimental values are in good agreement with this prediction
\beq{ratvpexp}
\frac {BR(B^+\rightarrow K^{*+} \eta')}{BR(B^+\rightarrow K^{*+} \eta)}
=\frac{4.9\pm 2.1}{19.3\pm 1.6} = 0.25 \pm 0.11  
\end{equation}
\beq{ratvpexpo}
\frac {BR(B^o\rightarrow K^{*o} \eta')}{BR(B^o\rightarrow K^{*o} \eta)}
=\frac{3.8 \pm 1.2}{15.9 \pm 1} = 0.24 \pm 0.08 
\end{equation}

\subsection{A bad relation for the sum $BR(B^+\rightarrow K^{*+} \eta')
+ BR(B^+\rightarrow K^{*+} \eta)$ 
independent of form factor differences}

We now examine more carefully the decays to the final states 
$K^{*+} \eta $ and $K^{*+} \eta'$ where significant violations of the simplest
model occur. To pinpoint these violations we choose
relations that are independent of form factor differences and do not depend on
the relation (\ref{formfac}). We 
assume that the penguin contribution to B decays into two charmless strange 
mesons denoted by $M_1$ and $M_2$ goes via the transition

\beq{pentrans2}
\ket{B^+} \rightarrow \ket{(\bar s q_g)_{W}; (\bar q_g u)_{S}} 
\rightarrow \ket{M_{1}; M_{2}}
\end {equation}
where $q_g$ and $\bar q_g$ denote a quark of pair of any flavor. The subscript
$g$ denotes that they come from the same gluon.  The subscript $W$ denotes that
the pair contains a $\bar s$ antiquark produced at the weak vertex and the
subscript $S$ denotes that the pair contains the spectator $u$ quark.  We
assume that the transition is flavor independent but the hadronization form
factors for weak and spectator vertices can be different. The transition matrix
element therefore satisfies the  relation, 

\beq{pengop2}
\bra{(\bar s u_g)_{W}; (\bar u_g u)_{S}}T_P\ket{B^+}=
\bra{(\bar s d_g)_{W}; (\bar d_g u)_{S}}T_P\ket{B^+}=
\bra{(\bar s s_g)_{W}; (\bar s_g u)_{S}}T_P\ket{B^+} 
\end {equation}

Since the spectator $u$ quark remains in all transitions, the transition that
would require a spectator $d$ quark vanishes. Thus the states $K^{*+}\eta_n$ and 
$K^{*+}\pi^o$ are both produced equally via the $K^{*+}(u\bar u) $ state
\beq{etapi}
BR(B^+\rightarrow K^{*+} \eta_n)= BR(B^+\rightarrow K^{*+} \pi^o)
\end {equation}
where we neglect phase space corrections.

The $B^+\rightarrow K^{*+} \eta_s$ and $B^+\rightarrow K^o \rho^+$
transitions both have final states with a vector meson containing the spectator
quark and a pseudoscalar containing the strange antiquark from the weak 
$\bar b \rightarrow \bar s$ transitions. The form factors are the same and 
\beq{etas}
BR(B^+\rightarrow K^{*+} \eta_s)= BR(B^+\rightarrow K^o \rho^+)
\end {equation}
Combining equations (\ref{etapi}) and (\ref{etas}) gives a sum rule independent
of the standard model mixing angle which can be compared with experiment.
\beq{sumruleVP}
\begin{array}{ccccccc}
BR(B^+\rightarrow K^{*+} \eta_n)+BR(B^+\rightarrow K^{*+} \eta_s) &=&
BR(B^+\rightarrow K^o \rho^+) + BR(B^+\rightarrow K^{*+} \pi^o)
\\
BR(B^+\rightarrow K^{*+} \eta)+ BR(B^+\rightarrow K^{*+} \eta') &=& 
BR(B^+\rightarrow K^o \rho^+) + BR(B^+\rightarrow K^{*+} \pi^o)
\\  \rm{BaBar ~ Data} & &
\\
(18.9 \pm 1.8 \pm 1.3) + (4.9 \pm 1.9 \pm 1.8) = 23.8 \pm X &= &
(8.0\pm 1.4 \pm 0.5) + (6.9 \pm 2.0 \pm 1.3) = 14.9 \pm X
\\  \rm{FFAG-Avg} & &
\\
(19.3 \pm 1.6) + (4.9 \pm 2.1) = 24.2 \pm X &= &
(8.0\pm 1.5) + (6.9 \pm 2.3) = 14.9 \pm X

\end{array}
\end {equation}

This sum rule is seen to be seriously violated in the same way as the previous
pseudoscalar
sum rule\cite{bkpfsifin} indicating an additional contribution to the  $\eta -
\eta'$ system.  
\beq{sumrulepsbr} 
BR(B^\pm \rightarrow K^\pm \eta') + BR(B^\pm
\rightarrow K^\pm \eta) \leq  BR(B^\pm \rightarrow K^\pm \pi^o) +  BR(B^\pm
\rightarrow \tilde K^o \pi^\pm) 
\eeq 
where $\tilde K^o$ denotes $ K^o$ for the
$B^+$ decay and $\bar K^o$ for the $B^-$ decay. 
The experimental
values\cite{HFAG} in units of $10^{-6}$ are 
\beq{sumrulepsxbr} 
BR( K^\pm \eta')(69.7 \pm 3.8) + BR( K^\pm \eta)  (2.2 \pm 0.3) \leq  
BR( K^\pm \pi^o ) (12.8 \pm 0.6) +  BR( K^o \pi^\pm) (23.1 \pm 1.0) 
\eeq

Whether this additional contribution arises from an electroweak penguin 
contribution\cite{chiang} or a difference in the wave functions from the
standard quark model is an open challenge for QCD. The possibility of adding
``intrinsic charm" to the wave functions\cite{Hareta,halzhit,atson,brodgard}
would mix an $\eta_c$ into the $\eta$ and/or $\eta'$ wave function.

The branching ratio $BR(B^+\rightarrow K^+ \eta_c)$ is  $9.1 \pm 1.3 \times
10^{-4}$ which is larger by a factor of 38 than the charmless branching ratio
$BR(B^+\rightarrow K^o \pi^+)$,  $24.1 \times 10^{-6}$. The difference in phase
space indicates an even larger ratio of the squares of the transition matrix
elements. Thus an admixture of only a few per cent of  $\eta_c$ into the $\eta$
and/or $\eta'$ wave function could eliminate the disagreements with the sum
rules. It would be of interest to find an experimental test which would
distinguish between an electroweak penguin contribution and between intrinsic
charm or $\eta_c$ admixture in the wave functions. Most tests in standard $\eta
- \eta'$ spectroscopy are not sensitive enough to detect these small
admixtures, but EWP contributions mat be related to other observables.

\section{A symmetry approach to charmless strange $B^+$ vector-pseudoscalar
decays}

\subsection{Exotic and nonexotic final states}

We now search for an alternative symmetry approach that can distinguish between
relations that neglect and exhibit the form factor differences. 

The penguin diagram for the decay of a pseudoscalar  $B^+$ meson to a charmless
strange vector pseudoscalar state has a final quark-antiquark-gluon state of
odd parity with the flavor quantum numbers of a kaon. It is the strange member
of a pseudoscalar SU(3) flavor octet but has two possible eigenvalues for the
generalized charge conjugation operator, also known as ``isoparity", which
defines the relative phases of the charge conjugate states in the same octet
and the eigenvalue under charge conjugation of its C-eigenstate members.
Dothan\cite{dothan}   generalized the idea of G parity from SU(2) to SU(3).
This developed further\cite{larry} on generalizations of isoparity. For weak
interactions, the generalizations of G  parity from SU(2) to SU(n) is not
directly relevant because charge conjugation  is badly violated like parity in
weak interactions. However one can multiply G  parity by normal space-inversion
parity to make GP. 

We call the generalization of GP to SU(3) and SU(n)  ``Dothan parity" and
investigate its relevance to weak decays, We first consider the V- spin (us)
subgroup of flavor SU(3) and the $G_V$ parity. The ordinary $K^+$ charged strange
pseudoscalar meson is a member of an octet whose charge-conjugate state, the
$\pi^o$ is even under C.  The pion isotriplet has isospin one and odd G parity.
The $K^+$ has V spin 1 and odd $G_V$ parity.  An ``exotic" pseudoscalar octet
can be defined whose nonstrange isovector is odd under charge conjugation and
whose positively charged strange member has V spin 1 and even $G_V$ parity.
This state is called exotic because it cannot be made from a quark-antiquark
pair. However, it can be made from a quark-antiquark pair and a gluon and can a
priori occur in a penguin diagram. Since the parity of the state is negative,
the normal state has even $G_VP$ like the pion has even $GP$; the exotic state
has odd $G_VP$ and its nonstrange isovector has odd $GP$ opposite to that of the
pion. 

Since the relations (\ref{nonexot}) are obtainable by using SU(3) flavor
symmetry to relate the decays of a high mass kaon, we see that these relations
assume that the initial state is ``normal" and that contributions from an
exotic initial state are negelected. Since GP is simply related to CP, one
can look for a possibility of identifying the normal state with a CP-conserving
transition and the exotic state with a CP-violating transition. 

For a simple case we first examine the $B_s \rightarrow \phi \eta$ decay. The
$B_s (\bar b s)$ is a pseudoscalar meson with odd parity. If it is a member of
any  symmetry multiplet which includes $\bar b b$ and $\bar s s$ pseudoscalar
states; e.g. an SU(2) symmetry in the $bs$ flavor space, the multiplet is even
under charge conjugation and therefore odd under CP. The $\phi \eta$ p-wave
state which can be produced in $B_s (\bar b s)$ decay is also odd under parity
but odd under charge conjugation and even under CP. Thus if this generalized CP
is conserved the decay $B_s \rightarrow \phi \eta$ is forbidden. Any symmetry
which contains both the $\bar b b$ and $\bar s s$  pseudoscalar states is badly
broken by mass differences. However we first consider the implications of such
symmetries for weak decays and leave symmetry breaking for later analysis.

We note that the final states $K^{*+}K^-$ and $K^{*-}K^+$ go into one another
under CP and similarly for $K^{*o}\bar K^o$ and $\bar K^{*o}K^o$.
Thus conservation of this generalized CP predicts
\beq{bssympred}
\begin{array}{ccccccc}
BR(B_s\rightarrow \phi \eta) &=& BR(B_s\rightarrow \phi \eta') = 0
\\
BR(B_s\rightarrow K^{*o} \bar K^o) &=& BR(B_s\rightarrow \bar K^{*o} K^o) 
\\
BR(B_s\rightarrow K^{*+} K^-) &=& BR(B_s\rightarrow K^{*-} K^+)
\end{array}
\end {equation}
Note that in the dominant penguin diagram 
\beq{pspeng}
B_s (\bar bs)\rightarrow (\bar s G s) \rightarrow (\bar su) (\bar us) 
\rightarrow K^+ K^-        
\end{equation}
where $K$ here denotes either pseudoscalar or vector meson,
the positive kaon is produced by combining the $u$ with the $\bar s$ produced
in the weak interaction; the negative kaon is produced by combining the $\bar
u$ with  the spectator $s$ quark. The form factors for producing vector and
pseudoscalar kaons from the two vertices are not expected to be equal. So the
equality and the selection rules are nontrivial and their conformation or
violation is interesting.

We now consider a $(ubs)$ SU(3) flavor symmetry which is the analog of the
usual $(uds)$ symmetry with the $d$ replaced by the $b$. Since there are no $d$
quarks in the $B^+$ or $K^+$ we can consider the weak decay

\beq{bkg}
      B^+\rightarrow K^+ G        
\end{equation}
as a transition between two octet states in the $(ubs)$ classification.  We
can also define the phases of the CKM matrix for the weak interaction in
the standard model with all CKM phases real in the $(ubs)$ flavor subspace.
There is therefore no $CP$ violation in this flavor subspace.

      If we now combine CP with flavor in this subspace to make Dothan parity,
we find that neutral members of the flavor octets on both sides of eq.
(\ref{bkg}) which are eigenstates of CP must have the same eigenvalue. This
tells us that the $K^+ G$ state must have the same $G_V$ parity as the kaon and
that the exotic state is forbidden.

       The relevant SU(3) coupling in conventional (uds) SU(3) is a (VPK)
coupling of three octets which is unique since the strong interaction
conserves charge conjugation. It is therefore also the same as the $K\pi$,
$K\eta$ and $K\eta'$ couplings to the $K^*(890)$.  

The experimental consequences of the extreme assumption (\ref{nonexot})  is
equivalent on one hand to the neglect of the differences between vector and
pseudoscalar form factors or on the other hand is equivalent to considering
only SU(3) octet final states with normal generalized charge conjugation and
neglecting the exotic contribution. 

\subsection{Possible dynamical justification for neglecting exotic final states}

Most conventional calculations of penguin diagrams do not consider the
classification of final states as exotic nor nonexotic. To define more
precisely the differences between the two approaches  we first write a general
expression for the transition matrix element of a penguin diagram. The
charmless strange $B^+$ decay is a weak transition denoted by an operator $T_W$ to an
intermediate state  $\ket{\bar s u G}$ followed by a strong transition
denoted by an operator $T_S$ to the final state.

\beq{pengopgen}
\bra {f}T_P\ket{B^+}=
\bra {f}T_S\ket{\bar s u G}\cdot\bra{\bar s u G}T_W\ket{B^+}
\end {equation}
The strong transition can be written as 
\beq{pengopgenstr}
\bra {f}T_S\ket{\bar s u G}=
\sum_i\bra {f}{H_i^{norm}}\rangle\cdot 
\bra {H_i^{norm}}T_S\ket{\bar s u G}
+ \sum_i\bra {f}{H_i^{ex}}\rangle\cdot 
 \bra {H_i^{ex}}T_S\ket{\bar s u G}
\end {equation}
where the sum over a complete set of strong interaction eigenstates is separated
into a sum over ``normal" states having odd $G_V$ denoted by
 $\ket {H_i^{norm}}$ and
a sum over exotic states having even $G_V$ denoted by $\ket{H_i^{ex}}$.
Since $G_V$ is conserved by strong interactions that conserve flavor SU(3),
the two sums are independent and there is no mixing between the states in the
two summations.

    We now see that if the strong interaction conserves flavor SU(3) symmetry, 
the symmetry alone places serious constraints on the observable branching
ratios. These are completely independent of detailed dynamical assumptions like
heavy quark symmetry, helicity conservation or factorization. 

The final two meson state is a linear combination of the contributions from the
two summations in eq. (\ref{pengopgenstr}), the even $G_V$ summation and the
odd $G_V$ summation. The branching ratios depend upon the relative magnitude
and phase of the two summations which are determined by unknown QCD dynamics. A
very small QCD interaction normally neglected will act differently on the two
summations if they are both appreciable and destroy any relative magnitude and
phase coherence.  Assuming that only the odd $G_V$  summation contributes leads
to our results where the amplitude with the quantum numbers of the kaon gives
branching ratios determined from V spin like those from decay of a high mass
kaon.

Feasible calculations involve choosing a particular set of intermediate states
in eq. (\ref{pengopgenstr}) and neglecting the contributions of others. Most
common calculations; e.g. refs. \cite{grorosbvp,chiang,chiangros} consider only
states of two quark-antiquark pairs and neglect multiparticle intermediate
states. The symmetry approach considered here neglects all intermediate states 
$\ket{H_i^{ex}}$ having exotic flavor quantum numbers but includes all
multiparticle states having normal quantum numbers. This corresponds to the
dual-resonance-model\cite{venez} approach in which the strong interaction $S$
matrix is represented by the sum of $S$ channel resonances which all have
non-exotic quantum numbers. 

It is difficult at this point to decide which approach is confirmed by rigorous
QCD, neglect of multiparticle intermediate states or neglect of exotic
intermediate states. However, the calculation considering  only
states of two quark-antiquark pairs generally do not analyze the $G_V$ parity 
of their expressions. If a $G_V$ analysis of their final states involve 
appreciable contributions having both $G_V$ parities, their predictions can be
destroyed by any neglected small QCD interaction.

QCD factorization is natural in the standard tree diagram where a $\bar b$
antiquark emits a high momentum $W$ which leaves the spectator quark before it
hadronizes into a meson and has no further strong interactions. The penguin
diagram is very different as the gluon and $\bar s$ are emitted in opposite
directions in the rest system of the spectator quark. The gluon must then
interact with both others to produce the final state. The assumption that the
gluon first produces a $q \bar q$ pair before interaction with the others can
be questioned. At this stage one can compare the experimental consequences of
both approaches.

\section{Further experimental tests of the Nonexotic model}

We now examine further experimental tests of the nonexotic model. The
assumption that charmless strange B decays are dominated by transitions via 
intermediate states with nonexotic quantum numbers can immediately be tested in
all final states. Only even values of $G_V P$ contribute; i.e. odd $G_V$ for 
odd parity final states and even $G_V$ for even parity states. This immediately
leads to the analog for vector-vector final states of the prediction
(\ref{bkrpred}), which is seen to agree with experiment.
\beq{bkrpredvv}
\begin{array}{ccccccc}
BR(B^+\rightarrow \rho^+ K^{*o}) &= BR(B^+\rightarrow \phi K^{*o}) 
\\
= 9.2\pm 1.5  &=  9.7 \pm 1.5
\end{array}
\end {equation}
where we have used the data from  the 
HFAG Average\cite{HFAG}.
Further data will be able to test the nonexotic model by 
the relations from the full generalization to the other cases of 
the vector pseudoscalar relations  (\ref{nonexot}). For the vector-vector case, 
\beq{nonexotvv}
\begin{array}{ccccccc}
2BR(B^+\rightarrow K^{*+} \omega) &=2BR(B^+\rightarrow K^{*+} \rho^o) &=
BR(B^+\rightarrow K^{*o} \rho^+) &=
BR(B^+\rightarrow \phi K^+) 
\\  &= &= 9.2\pm 1.5  &=
 9.7 \pm 1.5
\end{array}
\end {equation}
Analogous predictions can be made for other final states; e.g. the tensor-
pseudoscalar case including the 
$K^*_2(1430)$ tensor resonance which also has even parity and will be
dominated by even $G_V$ states.  
There is also the prediction  for the TP final
states including the $\eta$ and $\eta'$.
\beq{branvv}
\frac {BR(B^+\rightarrow K^{*+}_2 \eta)}{BR(B^+\rightarrow K^{*+}_2 \eta')}
=\frac {BR(B^o\rightarrow K^{*o}_2 \eta)}{BR(B^o\rightarrow K^{*o}_2 \eta')}
=\frac {BR(B^+\rightarrow K^+ \eta)}{BR(B^+\rightarrow K^+ \eta')}
=\frac {BR(B^o\rightarrow K^o \eta)}{BR(B^o\rightarrow K^o \eta')}
\end{equation}

\section*{Acknowledgements}

This research was supported in part by the U.S. Department of Energy, Division
of High Energy Physics, Contract DE-AC02-06CH11357. It is a pleasure to thank
Stanley Brodsky, Susan Gardner, Michael Gronau, Yuval Grossman, Marek Karliner,
Zoltan Ligeti, Yosef Nir, Jonathan Rosner, J.G. Smith, Frank Wuerthwein and 
Jure Zupan for discussions and comments.

%
\catcode`\@=11 
\def\references{
\ifpreprintsty \vskip 10ex
%
\hbox to\hsize{\hss \large \refname \hss }\else
\vskip 24pt \hrule width\hsize \relax \vskip 1.6cm \fi \list
{\@biblabel {\arabic {enumiv}}}
{\labelwidth \WidestRefLabelThusFar \labelsep 4pt \leftmargin \labelwidth
\advance \leftmargin \labelsep \ifdim \baselinestretch pt>1 pt
\parsep 4pt\relax \else \parsep 0pt\relax \fi \itemsep \parsep \usecounter
{enumiv}\let \p@enumiv \@empty \def \theenumiv {\arabic {enumiv}}}
\let \newblock \relax \sloppy
 \clubpenalty 4000\widowpenalty 4000 \sfcode `\.=1000\relax \ifpreprintsty
\else \small \fi}
\catcode`\@=12 

\end{document}